\documentstyle[preprint,aps]{revtex}
\preprint{USM-TH-150}
\begin{document}
\title{Coulomb interaction from the interplay between Confinement
and Screening}
\author{P. Gaete$^{1}$\thanks{
E-mail: patricio.gaete@fis.utfsm.cl} and E. I. Guendelman
$^{2}$\thanks{ E-mail: guendel@bgumail.bgu.ac.il}}
\address{$^1$Departamento de F\'{\i}sica, Universidad T\'{e}cnica
F. Santa Mar\'{\i}a, Casilla 110-V, Valpara\'{\i}so, Chile \\
$^2$Physics Department, Ben Gurion University, Beer Sheva 84105,
Israel} \maketitle
\begin{abstract}
It has been noticed that confinement effects can be described by
the addition of a  $ \sqrt { - F_{\mu \nu }^a F^{a\mu \nu } } $
term in the Lagrangian density. We now study the combined effect
of such "confinement term" and that of a mass term. The surprising
result is that the interplay between these two terms gives rise to
a Coulomb interaction. Our picture has a certain correspondence
with the quasiconfinement picture described by Giles, Jaffe and de
Rujula for QCD with symmetry breaking.
\end{abstract}
\smallskip

PACS number(s): 11.10.Ef, 11.15.Kc

\section{Introduction}

It is well known that one of the long standing problems in physics
is understanding the confinement physics from first principles.
Hence the challenge is to develop analytical approaches which
provide valuable insight and theoretical guidance. According to
this viewpoint, an effective theory in which confining potentials
are obtained as a consequence of spontaneous symmetry breaking of
scale invariance has been developed\cite{GaeteG}. In particular,
it was shown that a such theory relies on a scale invariant
Lagrangian of the type \cite{GuendeG}
\begin{equation}
{\cal L} = \frac{1}{4}w^2  - \frac{1}{2}w\sqrt { - F_{\mu \nu }^a
F^{a\mu \nu } }, \label{Pro1}
\end{equation}
where $F_{\mu \nu }^a  = \partial _\mu  A_\nu ^a  - \partial _\nu
A_\mu ^a  + gf^{abc} A_\mu ^b A_\nu ^c$, and $w$ is not a
fundamental field but rather is a function of $4$-index field
strength, that is,
\begin{equation}
w = \varepsilon ^{\mu \nu \alpha \beta } \partial _\mu  A_{\nu
\alpha \beta }. \label{Pro2}
\end{equation}
The $A_{\nu\alpha\beta}$ equation of motion leads to
\begin{equation}
\varepsilon ^{\mu \nu \alpha \beta } \partial _\beta  \left( {w -
\sqrt { - F_{\gamma\delta}^a F^{a\gamma\delta} } } \right) = 0,
\label{Pro3}
\end{equation}
which is then integrated to
\begin{equation}
w = \sqrt { - F_{\mu \nu }^a F^{a\mu \nu } }  + M .\label{Pro4}
\end{equation}
It is easy to verify that the $A_a^{\mu}$ equation of motion leads
us to
\begin{equation}
\nabla _\mu  \left( {F^{a\mu \nu }  + M\frac{{F^{a\mu \nu }
}}{{\sqrt { - F_{\alpha\beta }^b F^{b\alpha\beta } } }}} \right) =
0 . \label{Pro5}
\end{equation}
It is worth stressing at this stage that the above equation can be
obtained from the effective Lagrangian
\begin{equation}
{\cal L}_{eff}  =  - \frac{1}{4}F_{\mu \nu }^a F^{a\mu \nu }  +
\frac{M}{2}\sqrt { - F_{\mu \nu }^a F^{a\mu \nu } }. \label{Pro6}
\end{equation}
Spherically symmetric solutions of Eqs.(\ref{Pro5}) display, even
in the Abelian case, a Coulomb piece and a confining part. Also,
the quantum theory calculation of the static energy between two
charges displays the same behavior\cite{GaeteG}. It is well known
that the square root part describes string like
solutions\cite{Nielsen,Amer}.

Within this framework the aim of the present Letter is to extend
further the previous analysis by considering the effect of a mass
term. To this end we will compute the static potential of this
theory. In fact, we will show that the static potential for the
new theory gives rise to an effective Coulomb interaction. We
recall in passing that the static potential between a heavy quark
and antiquark is a tool of considerable theoretical interest which
is expected to provide the foundation for understanding
confinement. According to our approach, the interaction potential
between two charges is obtained once a suitable identification of
the physical degrees of freedom is made. This methodology has been
used previously in many examples for studying features of
screening and confinement in gauge theories \cite{GaeteA,GaeteB}.

\section{The Interplay between Confinement and Mass terms}

Some time ago, Giles, Jaffe and de Rujula \cite{Giles} proposed
that in the presence of spontaneous breaking of gauge symmetry
confinement in $QCD$ may become an approximate effect and there
could be in this case high mass states of unconfined quarks and
gluons. Their analysis was done in the context of the $MIT$ bag
model\cite{MIT}.

Subsequently this research was criticized by Georgi \cite{Georgi},
who argued that the confinement properties of $QCD$ will present
an obstacle for the s.s.b. of gauge symmetry.

Here we want to show that even if s.s.b. of gauge symmetry is not
in question and that there is indeed a mass term induced in the
action, then the dynamics of a theory which is governed by a
confining term ( explained in the previous section ) and a mass
term presents highly unexpected features.

Let us study an effective action of the form
\begin{equation}
{\cal L}_{eff}  =  - \frac{1}{4}F_{\mu \nu }^a F^{a\mu \nu }  +
\frac{M}{2}\sqrt { - F_{\mu \nu }^a F^{a\mu \nu } }  + \frac{\mu
^2}{2} A_\mu ^a A^{a\mu }, \label{Pro7}
\end{equation}
and let us study for simplicity the Abelian case. Then, equation
for the spherically symmetric case is
\begin{equation}
\nabla  \cdot \left( {{\bf E} + \frac{M}{{\sqrt 2 }}\hat {\bf r}}
\right) = \mu^2\phi. \label{Pro8}
\end{equation}
Looking for static solutions where also we set ${\bf A}=0$, that
is, ${\bf E}=-\nabla\phi$, we find that Eq.(\ref{Pro8}) becomes
\begin{equation}
\frac{1}{r}\frac{{d^2 }}{{dr^2 }}\left( {r\phi } \right) -
\frac{M}{{\sqrt 2 }}\frac{1}{r} + \mu ^2 \phi  = 0, \label{Pro9}
\end{equation}
which for $\mu^2=0$, has as solution\cite{GaeteG}
\begin{equation}
\phi  = \frac{C}{r} + \frac{M}{{\sqrt 2 }}r, \label{Proo9}
\end{equation}
displaying a confinement (M) part and a Coulomb part. Notice that
for $\mu^2\ne0$ the nature of the solutions is totally different,
being of the form
\begin{equation}
\phi  = C\frac{{e^{ - \mu r} }}{r} + \left( {\frac{M}{{\sqrt 2 \mu
^2 }}} \right)\frac{1}{r}. \label{Pro10}
\end{equation}
From Eq.(\ref{Pro10}) we can appreciate the interesting phenomenon
of the appearance of an effective Coulomb term, which depends on
both the confining term ($M$ dependence) and on the screening or
mass term ($\mu^2$ dependence). The confining term in
Eq.(\ref{Proo9}) has disappeared and is being replaced by a
Coulomb term, even for $\mu$ arbitrarily small. As
$\mu^2\rightarrow0$ instead of confinement one has an arbitrarily
strong Coulomb term. These general arguments can be put in a more
solid ground by the use of the full quantum mechanical
gauge-invariant variables formalism.

\section{Interaction energy}

As already mentioned, our immediate objective is to compute
explicitly the interaction energy between static pointlike sources
for the mode under consideration. The starting point is the
two-dimensional space-time Lagrangian obtained from (\ref{Pro7})
in the Abelian case and considering only $r,t$ dependence, a sort
of minisuper-space approach\cite{Benguria}.
\begin{equation}
{\cal L} = 4\pi r^2 \left\{ { - \frac{1}{4}F_{\mu \nu } F^{\mu \nu
} - \frac{M}{2 \sqrt {2}}\varepsilon _{\mu \nu } F^{\mu \nu } +
\frac{{\mu ^2 }}{2}A_\mu A^\mu  } \right\} - A_0 J^0 ,
\label{poten1}
\end{equation}
where $J^0$ is the external current, and $\mu$ is the mass for the
gauge fields. Here $\mu,\nu=0,1$, where $ x^1 \equiv r \equiv|{\bf
x}|$ and $\varepsilon^{01}=1$. We have used that in a two
dimensional space $(t,r)$, $\sqrt { - F^{\mu \nu } F_{\mu \nu } }
= \frac{{\varepsilon _{\mu \nu } F^{\mu \nu } }}{{\sqrt 2 }}$. It
is worthwhile sketching at this point the canonical quantization
of this theory from the Hamiltonian analysis point of view. The
canonical momenta read $\Pi ^\mu = - 4\pi x^2 \left( {F^{0\mu } +
\frac{M}{\sqrt {2}} \varepsilon ^{0\mu } } \right)$, which results
in the usual primary constraint $\Pi^0=0$, and $\Pi ^i = - 4\pi
x^2 \left( {F^{0i}  + \frac{M}{\sqrt {2}} \varepsilon ^{0i} }
\right)$. The canonical Hamiltonian following from the above
Lagrangian is:
\begin{equation}
H_0  = \int {dx} \left\{ {\Pi _1 \partial ^1 A^0  - \frac{1}{{8\pi
x^2 }}\Pi _1 \Pi ^1  - \frac{M}{\sqrt {2}}\varepsilon ^{01} \Pi _1
+ \pi x^2 M^2 - 2\pi x^2 \mu ^2 \left( {A_0 A^0  + A_1 A^1 }
\right) + A_0 J^0 } \right\}. \label{poten2}
\end{equation}
Requiring the primary constraint $\Pi _0=0$ to be preserved in
time yields the following secondary constraint
\begin{equation}
\Gamma \left( x \right) \equiv \partial _1 \Pi ^1  + 4\pi x^2 \mu
^2 A^0  - J^0  = 0. \label{potentia2}
\end{equation}
It is straightforward to see that both constraints are second
class. Thus, in order to convert the second class system into
first class we adopt the procedure described in
Refs.\cite{Clovis,Rabin}. In such a case we enlarge the original
phase space by introducing a canonical pair of fields $\theta$ and
$\Pi_\theta$. Then a new set of first class constraints can be
defined in this extended space:
\begin{equation}
\Lambda _1  \equiv \Pi _0  + 4\pi x^2 \mu ^2 \theta  = 0,
\label{vin1}
\end{equation}
and
\begin{equation}
\Lambda _2  \equiv \Gamma  + \Pi _\theta   = 0. \label{vin2}
\end{equation}
It is easy to verify that the new constraints are first class.
Therefore the new effective Lagrangian reads
\begin{equation}
 {\cal L} = 4\pi r^2 \left\{ { - \frac{1}{4}F_{\mu \nu } \left( {1 +
\frac{{\mu ^2 }}{\Box}} \right)F^{\mu \nu }  -
\frac{M}{2\sqrt{2}}\varepsilon _{\mu \nu } F^{\mu \nu } } \right\}
- A_0 J^0. \label{Mas1}
\end{equation}
We now restrict our attention to the Hamiltonian framework of this
theory. The canonical momenta read $\Pi ^\mu   =  - 4\pi x^2
\left[ {\left( {1 + \frac{{\mu ^2 }}{\Box }} \right)F^{0\mu } +
\frac{M}{\sqrt{2}}\varepsilon ^{0\mu } } \right]$. This yields the
usual primary constraint $\Pi^0=0$, and $\Pi ^i   =  - 4\pi x^2
\left[ {\left( {1 + \frac{{\mu ^2 }}{\Box }} \right)F^{0i } +
\frac{M}{\sqrt{2}}\varepsilon ^{0i } } \right]$. Therefore the
canonical Hamiltonian takes the form
\begin{equation}
\begin{array}{r}
H_C  = \int {dx} \left\{ { - A_0 \left( {\partial _1 \Pi ^1  - J^0
} \right) - \frac{1}{{8\pi x^2 }}\Pi _1 \left( {1 + \frac{{\mu ^2
}}{ \Box }} \right)^{ - 1} \Pi ^1  - \frac{M}{\sqrt{2}}\left( {1 +
\frac{{\mu ^2
}}{\Box }} \right)^{ - 1} \varepsilon ^{01} \Pi _1 } \right\} + \\
+ \int {dx\left\{ {\pi M^2 \left( {1 + \frac{{\mu ^2 }}{\Box}}
\right)^{ - 1} } x^2 \right\}}. \label{Mas2}
\end{array}
\end{equation}
Temporal conservation of the primary constraint $\Pi_0$ leads to
the secondary constraint $\Gamma _1 \left( x \right) \equiv
\partial _1 \Pi ^1  - J^0  = 0$. It is straightforward to check that
there are no further constraints in the theory. The extended
Hamiltonian that generates translations in time then reads $H =
H_C  + \int d x \left( {c_0 (x)\Pi_0 (x) + c_1 (x)\Gamma _1 (x)}
\right)$, where $c_0(x)$ and $c_1(x)$ are the Lagrange
multipliers. Moreover, it follows from this Hamiltonian that $
\dot{A}_0 \left( x \right) = \left[ {A_0 \left( x \right),H}
\right] = c_0 \left( x \right)$, which is an arbitrary function.
Since $\Pi_0 = 0$, neither $A^0$ nor $\Pi^0$ are of interest in
describing the system and may be discarded from the theory. As a
result, the Hamiltonian becomes
\begin{equation}
H = \int {dx} \left\{ { - \frac{1}{{8\pi x^2 }}\Pi _1 \left( {1 +
\frac{{\mu ^2 }}{\Box }} \right)^{ - 1} \Pi ^1  -
\frac{M}{\sqrt{2}}\left( {1 + \frac{{\mu ^2 }}{\Box }} \right)^{ -
1} \varepsilon ^{01} \Pi _{01} + c^ \prime  \left( {\partial _1
\Pi ^1  - J^0 } \right)} \right\}, \label{Mas3}
\end{equation}
where $c^ \prime  \left( x \right) = c_1 \left( x \right) - A_0
\left( x \right)$.

According to the usual procedure we introduce a supplementary
condition on the vector potential such that the full set of
constraints becomes second class. A convenient choice is found to
be \cite{GaeteA,GaeteB,GaeteG}
\begin{equation}
\Gamma _2 \left( x \right) \equiv \int\limits_{C_{\xi x} } {dz^\nu
} A_\nu \left( z \right) \equiv \int\limits_0^1 {d\lambda x^1 }
A_1 \left( {\lambda x} \right) = 0, \label{Mas4}
\end{equation}
where  $\lambda$ $(0\leq \lambda\leq1)$ is the parameter
describing the spacelike straight path $ x^1  = \xi ^1  + \lambda
\left( {x - \xi } \right)^1 $, and $ \xi $ is a fixed point
(reference point). There is no essential loss of generality if we
restrict our considerations to $ \xi ^1=0 $. In this case, the
only nontrivial Dirac bracket is
\begin{equation}
\left\{ {A_1 \left( x \right),\Pi ^1 \left( y \right)} \right\}^ *
= \delta ^{\left( 1 \right)} \left( {x - y} \right) -
\partial _1^x \int\limits_0^1 {d\lambda x^1 } \delta ^{\left( 1
\right)} \left( {\lambda x - y} \right). \label{Mas5}
\end{equation}

We are now equipped to compute the interaction energy between
pointlike sources in the model under consideration, where a
fermion is localized at the origin $ {\bf 0}$ and an antifermion
at $ {\bf y}$. In order to accomplish this purpose, we will
calculate the expectation value of the energy operator $H$ in the
physical state $ |\Phi\rangle$. From our above discussion, we see
that $\left\langle H \right\rangle _\Phi$ reads
\begin{equation}
\left\langle H \right\rangle _\Phi   = \left\langle \Phi
\right|\int {dx} \left( { - \frac{1}{{8\pi x^2 }}\Pi _1 \left( {1
+ \frac{{\mu ^2 }}{\Box }} \right)^{ - 1} \Pi ^1  -
\frac{M}{\sqrt{2}}\left( {1 + \frac{{\mu ^2 }}{\Box }} \right)^{ -
1} \varepsilon ^{01} \Pi _{01} } \right)\left| \Phi \right\rangle.
\label{Mas6}
\end{equation}
Since the fermions are taken to be infinitely massive (static), we
can substitute $\Box$ by $-\partial_1^2$ in Eq.(\ref{Mas6}). Here
$-\partial_1^2$ refers to the radial part of the spherical
Laplacian. In such a case we write
\begin{equation}
\left\langle H \right\rangle _\Phi   = \left\langle \Phi
\right|\int {dx} \left( { - \frac{1}{{8\pi x^2 }}\Pi _1 \left( {1
- \frac{{\mu ^2 }}{{\partial _1^2 }}} \right)^{ - 1} \Pi ^1  -
\frac{M}{\sqrt{2}}\left( {1 - \frac{{\mu ^2 }}{{\partial _1^2 }}}
\right)^{ - 1} \varepsilon ^{01} \Pi _1 } \right)\left| \Phi
\right\rangle. \label{Mas7}
\end{equation}
Next, as was first established by Dirac\cite{Dirac}, the physical
state can be written as
\begin{equation}
\left| \Phi  \right\rangle  \equiv \left| {\overline \Psi  \left(
\bf y \right)\Psi \left( \bf 0 \right)} \right\rangle  = \overline
\psi \left( \bf y \right)\exp \left( {ie\int\limits_{\bf 0}^{\bf
y} {dz^i } A_i \left( z \right)} \right)\psi \left(\bf 0
\right)\left| 0 \right\rangle, \label{Mas8}
\end{equation}
where $\left| 0 \right\rangle$ is the physical vacuum state and
the line integral appearing in the above expression is along a
spacelike path starting at $\bf 0$ and ending $\bf y$, on a fixed
time slice. From this we see that the fermion fields are now
dressed by a cloud of gauge fields.

Taking into account the above Hamiltonian structure, we observe
that
\begin{equation}
\Pi _1 \left( x \right)\left| {\overline \Psi  \left( y
\right)\Psi \left( 0 \right)} \right\rangle  = \overline \Psi
\left( y \right)\Psi \left( 0 \right)\Pi _1 \left( x \right)\left|
0 \right\rangle  - e\int_0^y {dz_1 } \delta ^{\left( 1 \right)}
\left( {z_1  - x} \right)\left| \Phi  \right\rangle. \label{Mas9}
\end{equation}
Inserting this back into (\ref{Mas7}), we get
\begin{equation}
\left\langle H \right\rangle _\Phi   = \left\langle H
\right\rangle _0  - \frac{{e^2 }}{{4\pi }}\frac{{e^{ - \mu L}
}}{L} - \frac{{Me}}{{\sqrt{2}4\pi \mu ^2 }}\frac{1}{L},
\label{Mas10}
\end{equation}
where $\left\langle H \right\rangle _0  = \left\langle 0
\right|H\left| 0 \right\rangle$ and with $|y|\equiv L$. Since the
potential is given by the term of the energy which depends on the
separation of the two fermions, from the expression (\ref{Mas10})
we obtain
\begin{equation}
V=- \frac{{e^2 }}{{4\pi }}\frac{{e^{ - \mu L} }}{L} -
\frac{{Me}}{{\sqrt{2}4\pi \mu ^2 }}\frac{1}{L}. \label{Mas11}
\end{equation}
In this way the static interaction between fermions arises only
because of the requirement that the $\left| {\overline \Psi \Psi }
\right\rangle$ states be gauge invariant.

\section{Final Remarks}

From our final expression for the heavy interquark potential we
see that:

a) For $\mu^2=0$ the theory describes an exactly confining phase.

b) For $\mu^2\ne0$ but $\mu^2$ very small, we observe that the
linear potential is now replaced by a Coulomb potential which is
however a very strong one. In this limit, states will be indeed
bound, that is, confined due to the very strong Coulomb potential
unless they correspond to very high excitations. Indeed the
"ionization energy" of this system goes to infinity as
$\mu^2\longrightarrow0$. However the Coulomb potential is not
exactly confining, therefore, even for small $\mu^2$, the
confining nature the potential is lost. In general, this picture
agrees qualitatively with that of Giles, Jaffe and de Rujula of
quasiconfinement for $QCD$ with a small gauge symmetry breaking
term\cite{Giles}.

\section{ACKNOWLEDGMENTS}

One of us (E.I.G.) wants to thank the Physics Department of the
Universidad T\'{e}cnica F. Santa Mar\'{\i}a for hospitality. P.G.
would like to thank I. Schmidt for his support.

\end{document}